\documentclass[aip,amsmath,amssymb,reprint,longbibliography]{revtex4-1}

\usepackage{graphicx}
\usepackage{dcolumn}
\usepackage{bm}
\usepackage[utf8]{inputenc}
\usepackage[T1]{fontenc}
\usepackage{mathptmx}
\usepackage{upgreek}
\usepackage[colorlinks,plainpages=false,linkcolor=blue,urlcolor=blue,citecolor=blue,pdfpagemode=UseNone,pdfstartview=FitH]{hyperref}

\begin{document}

\title[Internal Photo Effect from a Single Quantum Emitter]{Internal Photo Effect from a Single Quantum Emitter}

\author{P. Lochner}
\email{pia.lochner@uni-due.de}

\author{J. Kerski}
\affiliation{Faculty of Physics and CENIDE, University of Duisburg-Essen, 47057 Duisburg, Germany}

\author{A. Kurzmann}
\affiliation{Faculty of Physics and CENIDE, University of Duisburg-Essen, 47057 Duisburg, Germany}
\affiliation{Solid State Physics Laboratory, ETH Zurich, 8093 Zurich, Switzerland}

\author{A. D. Wieck}

\author{A. Ludwig}
\affiliation{Lehrstuhl für Angewandte Festk\"orperphysik, Ruhr-Universit\"at Bochum, 44780 Bochum, Germany}

\author{M. Geller}

\author{A. Lorke}
\affiliation{Faculty of Physics and CENIDE, University of Duisburg-Essen, 47057 Duisburg, Germany}

\date{\today}

\begin{abstract}
We demonstrate by time-resolved resonance fluorescence measurements on a single self-assembled quantum dot an internal photo-effect that emits electrons from the dot by an intra-band excitation. We find a linear dependence of the optically generated emission rate on the excitation intensity and use a rate equation model to deduce the involved rates. The emission rate is tunable over several orders of magnitude by adjusting the excitation intensity. Our findings show that a process that is well known in single atom spectroscopy (i.e. photo ionization) can also be observed in the solid state. The results also quantify an important, but mostly neglected, mechanism that may fundamentally limit the coherence times in solid-state quantum optical devices. 
\end{abstract}

\maketitle

The non-resonant excitation of electrons from a bound state into a continuum plays an important role in both basic and applied physics. The best-known example is the explanation of the photoelectric effect by Einstein\cite{Einstein1905}, which laid the groundwork for the development of quantum mechanics. In solid-state physics, a similar ‘internal’ photo effect, i.e. the non-resonant excitation of defect-bound charge carriers into the conduction or valence band continuum, is the basis of numerous photo detectors\cite{DiDomenico1964,Renker2009,Brenn1986}. Both of these non-resonant processes are accessible only through ensemble measurements. Resonant excitation, on the other hand, has been refined in both atomic and solid-state physics to enable studies on the energy structure and charge dynamics in single, confined quantum systems\cite{Kuhlmann2013,Haupt2014,Pinotsi2011,Kurzmann2016b,Kurzmann2019,Schulte2015,Fuchs2008,Batalov2009,Letokhov1985,Ma2017,Mueller2014,Ratschbacher2012}. In single atom spectroscopy, non-resonant electron excitation into the vacuum, called photo ionization, is a well studied effect that is generally considered detrimental, because it will destroy the object under investigation, i.e. the neutral atom in the magneto-optical trap\cite{Henkel2010}. Similar studies for solid-state single quantum emitters, however, have so far been missing.

Here, we report on the tunable bound-to-continuum excitation of electrons from a single confined system, specifically a single InAs quantum dot. The dot is embedded in a GaAs matrix, and the GaAs conduction band constitutes the continuum, into which the electron is excited. We use resonance fluorescence\cite{Vamivakas2009,Fallahi2010} with high spatial and energy resolution\cite{Matthiesen2012,Kuhlmann2015,Prechtel2013} to monitor the charging state of the quantum dot, while it is illuminated with pulsed non-resonant laser light. This way, we are able to observe the electron ejection as well as the re-occupation in a time-resolved n-shot measurement. We find that the emission rate is tunable over several orders of magnitude by adjusting the intensity of the non-resonant laser excitation. The repopulation of the ionized quantum dot, which takes place by tunneling of electrons into the dot from a charge reservoir, is found to be unaffected by the intensity of the non-resonant laser light. Model calculations confirm our picture of light-induced emission and the subsequent refilling of the quantum dot by tunneling. 

The measurements were performed on a sample grown by molecular beam epitaxy (MBE)\cite{Petroff2001}. A single layer of self-assembled In(Ga)As QDs is embedded in a p-i-n diode, where a highly p-doped GaAs layer on top defines an epitaxial gate. A degenerately n-doped GaAs layer acts as an electron reservoir, which is weakly coupled to the QDs with tunneling rates below 1\,ms\textsuperscript{-1} (see\cite{Lochner2019} for details about the sample). The QD states can energetically be shifted with respect to the Fermi energy in the electron reservoir by an applied gate voltage $V_\mathrm{G}$\cite{Hoegele2004}. This way, the charge state of the dots can be controlled by electron tunneling through the AlGaAs/GaAs barrier\cite{Geller2019}. In a confocal microscope setup within a bath cryostat at 4.2\,K, the sample is investigated by resonance fluorescence (RF, see\cite{Lochner2020} for details about setup and RF measurements). When the dot is in a neutral charging state, the RF signal corresponds to the creation and recombination of an electron-hole pair or exciton ($\mathrm{X}^0$). When the quantum dot is charged with a single electron, the RF excites the trion ($\mathrm{X}^-$), i.e. a bound state of two electrons and one hole. 

\begin{figure}
	\includegraphics{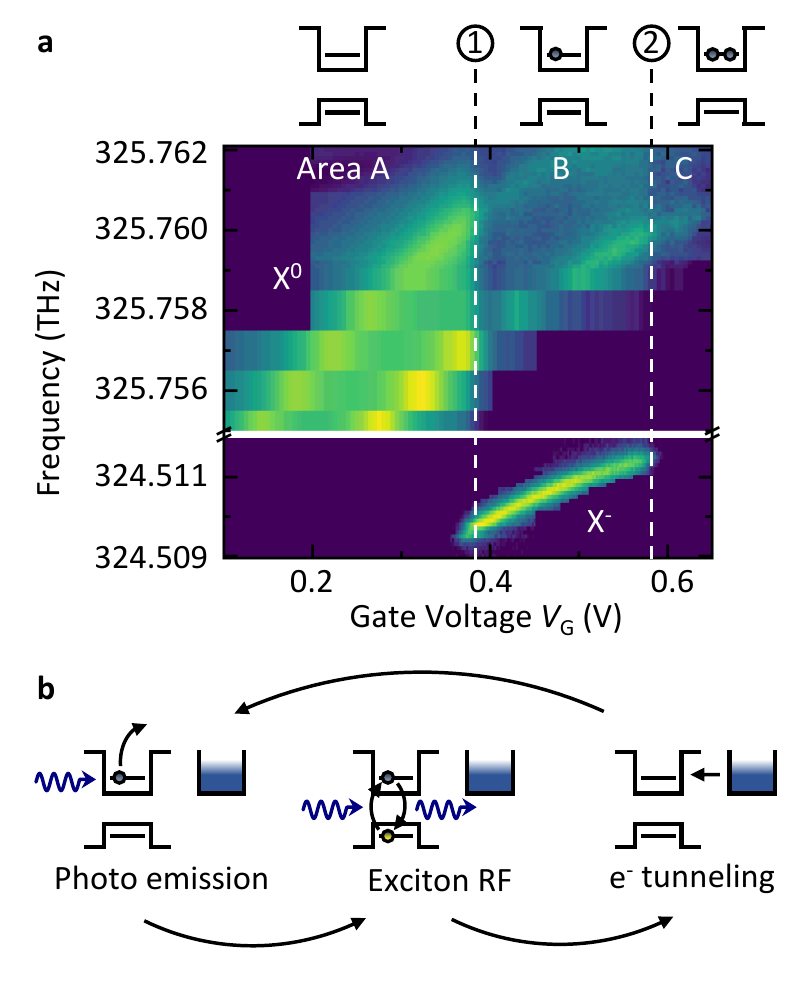}
	\caption{(a) Resonance fluorescence intensity of the exciton $\mathrm{X}^0$ and trion $\mathrm{X}^-$ transition at different gate voltages and laser excitation frequencies. Dashed lines mark the gate voltages where the first (at 1) and the second (at 2) electron tunnels into the QD, respectively. The $\mathrm{X}^-$ transition is observed at gate voltages $V_\mathrm{G}=0.38\,$V to $0.58\,$V and frequencies 324.5095\,THz to 324.5115\,THz where one electron occupies the QD. The $\mathrm{X}^0$ transition should be forbidden in this voltage range. (b) Schematic representation of photo emission which allows $\mathrm{X}^0$ transitions at high gate voltages. For gate voltages in area B, the photo emission removes the access electron (left figure) so that the exciton transition $\mathrm{X}^0$ is visible at frequencies 325.7575\,THz to 325.762\,THz (middle figure) until the QD is recharged from the reservoir (right figure).}
	\label{Fig1} 
\end{figure}

Figure~\ref{Fig1}(a) shows a map of the RF intensity of the exciton $\mathrm{X}^0$ and the trion $\mathrm{X}^-$ transition versus gate voltage $V_\mathrm{G}$ and laser frequency $\nu$. All measurements of the $\mathrm{X}^0$ ($\nu>325.755$\,THz) were performed at a laser intensity of $4\cdot10^{-2}\,\upmu\mathrm{W}/\upmu\mathrm{m}^2$. The measurements in the $\mathrm{X}^-$ region of area B ($\nu<324.512$\,THz) were performed at lower laser intensities ($7.9\cdot10^{-6}\,\upmu\mathrm{W}/\upmu\mathrm{m}^2$). The fine structure splitting of the exciton\cite{Gammon1996} and the upward shift with increasing gate voltage, caused by the quantum confined Stark effect\cite{Li2000}, are visible as expected. In area B ($V_\mathrm{G}=0.384$\,V to $V_\mathrm{G}=0.579$\,V), one electron occupies the QD, and the trion transition $\mathrm{X}^-$ is possible. At higher gate voltages in area C, a second electron tunnels into the QD and the trion transition quenches again. Surprisingly, the exciton transition $\mathrm{X}^0$ is visible even in area B of Figure~\ref{Fig1}, at gate voltages, where it should be forbidden, as one electron should constantly be present in the dot. This can be explained by the internal photo effect, which is schematically depicted in Figure~\ref{Fig1}(b). A laser photon interacts with the electron in the dot and the internal photo effect excites it into the conduction band (left panel in Figure~\ref{Fig1}(b)). This empties the quantum dot and makes it possible to drive the exciton transition $\mathrm{X}^0$ in the QD at a gate voltage, where it is usually forbidden (middle panel in Figure~\ref{Fig1}(b)). The exciton transition quenches again, when an electron tunnels into the dot from the reservoir (right panel in Figure~\ref{Fig1}(b)). 

\begin{figure*}
	\includegraphics{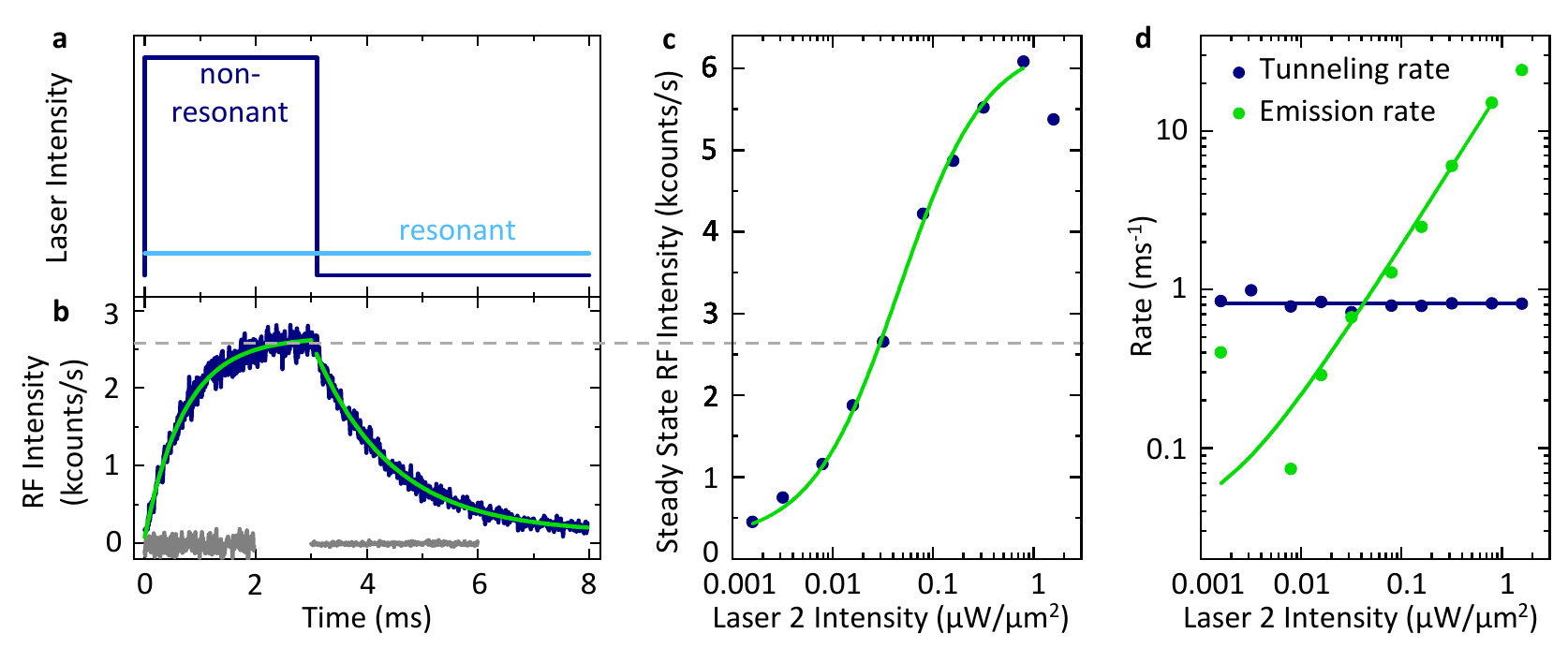}
	\caption{Time-resolved RF n-shot measurement of the internal photo effect. (a) Measurement scheme: One laser (bright blue) at constant, low intensity is tuned to the exciton resonance. A second laser (dark blue) is switched on for 3\,ms at a frequency of 324\,THz, which is energetically below all interband transitions. (b) Electron emission and tunneling transient at a laser intensity of $3.2\cdot10^{-2}$\,$\upmu\mathrm{W}/\upmu\mathrm{m}^2$ of the non-resonant laser: After the laser is switched on, an increasing RF emission from the $\mathrm{X}^0$ transition can be detected. After switching off the laser at 3\,ms, the $\mathrm{X}^0$ emission quenches again (dark blue line). Green lines are fits to the data using the rate equation given in the text (increasing part) and using a simple exponential decay function (decreasing part), respectively. The gray line shows that no RF counts are detected if the non-resonant laser is switched on without the resonant one. (c) The RF amplitude of the transient in (b) increase with increasing intensity of the non-resonant laser. The green line is a fit to the data using the rate equation for the limit of $t\rightarrow\infty$. (d) For increasing laser intensities of the non-resonant laser, the tunneling rates (dark blue) remain constant and the emission rates (green) increase linearly. Due to a constant offset, the straight does not look straight in the double logarithmic depiction.}
	\label{Fig2} 
\end{figure*}

To validate this model, two-color time-resolved RF measurements were performed. The gate voltage was set to 0.53\,V, which is in area B. Laser~1 is tuned into resonance with the $\mathrm{X}^0$ transition at this gate voltage ($\nu_\mathrm{\hspace{0.5mm}res}=325.7592\,\mathrm{THz}$). This laser is constantly switched on at a low intensity ($1.6\cdot10^{-3}$\,$\upmu\mathrm{W}/\upmu\mathrm{m}^2$, see light blue line in Figure~\ref{Fig2}(a)), so that it causes nearly no internal photo emission. Laser~2 is tuned out of resonance to $\nu_\mathrm{\hspace{0.5mm}non-res}=324\,\mathrm{THz}$, a frequency that is below all excitonic resonances, ruling out any band-to-band transitions. This second laser is switched on for 3\,ms at an intensity of $3.2\cdot10^{-2}$\,$\upmu\mathrm{W}/\upmu\mathrm{m}^2$, higher than the first resonant laser~1. Consequently, resonant laser~1 detects the empty state of the QD by driving the exciton transition $\mathrm{X}^0$, while the non-resonant laser~2 induces the photo effect. In short: No $\mathrm{X}^0$ photon means no photo effect has occurred. Figure~\ref{Fig2}(b) shows an n-shot pulsed measurement (dark blue line). After the non-resonant laser is switched on at $t=0$, the RF intensity increases until it saturates at $t=3\,$ms. This is governed by two processes: The internal photo emission, driven by the non-resonant laser, ionizes the dot with the rate $\gamma_\mathrm{\hspace{0.5mm}E}$ and enables exciton transition. At the same time, electrons from the reservoir can tunnel into the QD with the tunneling rate $\gamma_\mathrm{\hspace{0.5mm}In}$ and repopulate it. The RF intensity increases until a steady state between emission and tunneling is reached. After switching off the non-resonant laser, only the electron tunneling remains and quenches the QD transition, starting at $t=3\,$ms in Figure~\ref{Fig2}(b). Therefore, the decreasing RF signal after switching off the non-resonant laser can be fitted by a single exponential function (see green line for $t>3$\,s in Figure~\ref{Fig2}(b)). The increasing RF intensity in the QD ionization regime between $t=0$ and 3\,ms in Figure~\ref{Fig2}(b) can be described by a simple rate equation, where we use the probabilities that the QD is unoccupied ($P_0$) or occupied with one electron ($P_1$), with $P_0+P_1=1$. In a first step, we neglect laser~1 and look at the occupation dynamics induced by the non-resonant laser~2. Then, the probability of an unoccupied QD changes in time as $\dot{P_0}(t)=-\gamma_\mathrm{\hspace{0.5mm}In}P_0(t)+\gamma_\mathrm{\hspace{0.5mm}E}P_1(t)$. Together with the initial condition that one electron is charged in the QD, $P_0(0)=0$, one finds
\begin{equation}
	\label{P0}
	P_0(t)=\frac{\gamma_\mathrm{\hspace{0.5mm}E}}{\gamma_\mathrm{\hspace{0.5mm}In}+\gamma_\mathrm{\hspace{0.5mm}E}}\left(1-\mathrm{e}^{-\left(\gamma_\mathrm{\hspace{0.5mm}In}+\gamma_\mathrm{\hspace{0.5mm}E}\right)t}\right).
\end{equation}
The RF probing of the occupation state (by laser~1) is much faster than the dynamics of charging and discharging the dot. Therefore, the RF signal accurately reflects the occupation of the dot:  Whenever the dot is empty, the RF signal is "on", with an intensity $I_0$. When the dot is occupied, the fluorescence intensity vanishes, $I=0$ [see, Refs \onlinecite{Kurzmann2019,Lochner2020}].  
Consequently, the n-shot average of the RF intensity, $I(t)$, is directly proportional to the probability that the dot is empty
\begin{equation}
	\label{I}
	I(t)=I_0P_0(t).
\end{equation}
The tunneling rate $\gamma_\mathrm{\hspace{0.5mm}In}$ for electron tunneling from the reservoir into the QD can be obtained by a fit to the data in Figure~\ref{Fig2}(b), solid green line for $t>3\,$ms. The green line for $t<3\,$ms is a fit using the equations~(\ref{P0}) and (\ref{I}). In order to rule out the possibility that the observed increasing RF intensity is due to the non-resonant laser, this measurement is repeated with only the non-resonant laser~2 being switched on and without the resonant laser~1. The gray line in Figure~\ref{Fig2}(b) shows no RF intensity, thus, the observed RF transients indeed reflect the probability for the dot being empty, which is a further indication that here, the internal photo effect is observed. 

Figure~\ref{Fig2}(c) and (d) show the dependence of the internal photo effect on the laser intensity of the non-resonant laser~2. The rates are displayed in Figure~\ref{Fig2}(d). The tunneling rate remains constant at a value of $\gamma_\mathrm{\hspace{0.5mm}In}=(0.82\pm0.02)$\,ms\textsuperscript{-1} for all laser intensities of the non-resonant laser~2. This is in agreement with tunneling rates, which were measured on the same sample before\cite{Lochner2020}. At the same time, the photo emission rate increases linearly with increasing laser intensity of the non-resonant laser $I_\mathrm{\hspace{0.5mm}Laser2}$: $\gamma_\mathrm{\hspace{0.5mm}E}=(18.8\pm0.3)\,\frac{\mathrm{ms}^{-1}}{\upmu\mathrm{W}/\upmu\mathrm{m}^2}I_\mathrm{\hspace{0.5mm}Laser2}$ (see green fit in Figure~\ref{Fig2}(d); the data point at highest laser intensity ($1.6$\,$\upmu\mathrm{W}/\upmu\mathrm{m}^2$) of the non-resonant laser~2 is excluded from the fit. Here, the fluorescence intensity decreases again, due to photon-induced electron capture into the QD, see Kurzmann et al.\cite{Kurzmann2016a}). Also note that the influence of the constant illumination by the probe laser~1 only leads to a constant shift of the data points in Figure~\ref{Fig2}(d) (offset: $(18.8\pm0.3)\,\frac{\mathrm{ms}^{-1}}{\upmu\mathrm{W}/\upmu\mathrm{m}^2}I_\mathrm{\hspace{0.5mm}Laser1}$), so that the linear dependence is not affected. Thus, the emission rate is tunable over several orders of magnitude by adjusting the laser intensity. The linear dependence on the laser intensity supports the explanation of an internal photo emission. In Figure~\ref{Fig2}(c), the steady state RF intensity increases with increasing laser intensity: At higher laser intensity, the electron is emitted more often from the QD by the photo effect. The empty dot can be excited more often on the exciton transition and the RF intensity increases accordingly. The data points can be fitted by $I(t\rightarrow\infty)=I_0P_0(t\rightarrow\infty)$ (green line in Figure~\ref{Fig2}(c)).

We would like to point out that the weak coupling of the dot to the electron reservoir in our sample is the reason why the internal photo effect is so clearly observed in the present experiment. After the photo emission, the small tunneling rate leaves the dot empty, on average for ms, so that the intensity of the $\mathrm{X}^0$ suffices to detect an internal photo emission in a single semiconductor quantum dot.

\begin{figure}
	\includegraphics{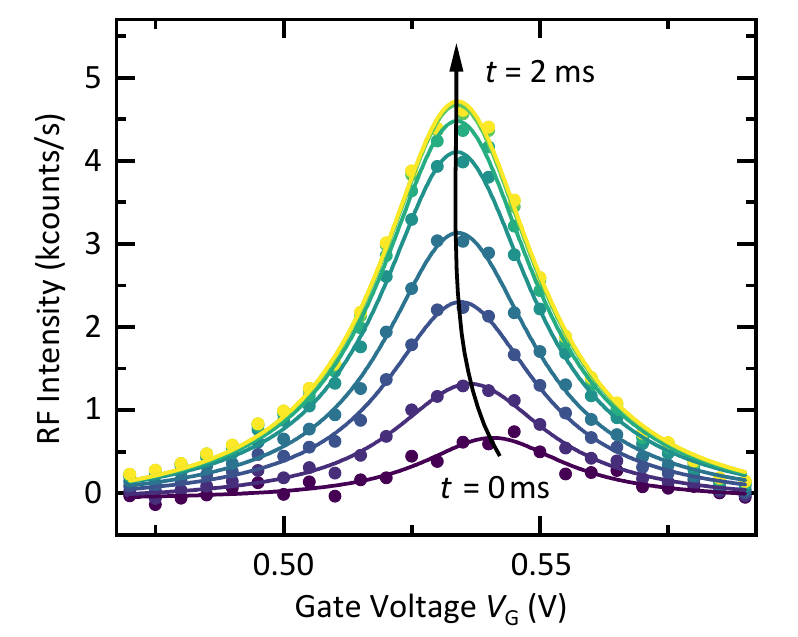}
	\caption{Gate voltage dependent RF intensity of the exciton transition after the non-resonant laser~2 is turned on for $t=0\,...\,2\,$ms. The intensity increases and the maximum shifts to lower gate voltages. At all times, the line shape is Lorentzian.}
	\label{Fig3}
\end{figure}

To answer the question, which effect the photo emitted electrons in the vicinity of the QD have on the QD emission, the evolution of the Lorentzian line shape of the exciton transition after switching on the non-resonant laser is investigated (Figure~\ref{Fig3}). The measurements were conducted like the measurements in Figure~\ref{Fig2}, with a non-resonant laser~2 intensity of $1.6\cdot10^{-1}$\,$\upmu\mathrm{W}/\upmu\mathrm{m}^2$. The $\mathrm{X}^0$ resonance was scanned by detuning the gate voltage from 0.47\,V to 0.59\,V. Cuts through the RF intensities at times $t=0\,...\,2\,$ms, after switching on the non-resonant laser~2, yield the results in Figure~\ref{Fig3}. With increasing time $t$, the RF intensity increases at all gate voltages, until it saturates. For all times, the $\mathrm{X}^0$ resonance is Lorentzian shaped (see fitted lines in Figure~\ref{Fig3}). The maximum of the resonance shifts to lower gate voltages with increasing time. In this sense, Figure~\ref{Fig3} shows an energetic shift of the exciton transition due to changes in the electrostatic environment of the dot. The photo emitted electrons are trapped in the vicinity of the dot and shift the exciton resonance due to the quantum confined Stark effect\cite{Arnold2014}. These results are in good agreement with measurements of Kurzmann et al.\cite{Kurzmann2016a}, where the capture of photo generated electrons into a quantum dot was shown.

In conclusion, we have observed the internal photo effect on a single semiconductor quantum dot. The electron emission rate was found to be tunable over several orders of magnitude, following a linear dependence on the incident laser intensity. We believe that our findings are of quite general relevance for all building blocks of quantum information technologies. Such devices are commonly described as ideal two-level systems. However, in most practical implementations, such as atoms, defects in solids, or semiconductor nanostructures, the confinement is limited, and non-resonant excitation into continuum states are possible. Photo-ionization and the internal photo effect demonstrated here are additional decoherence channels that limit the lifetimes of quantum states under optical excitation. Therefore, such processes should be taken into account when considering specific optical protocols for quantum information technologies.

\begin{acknowledgments}
This work was funded by the Deutsche Forschungsgemeinschaft (DFG, German Research Foundation) – Project-ID 278162697 – SFB 1242, and the individual research grant No. GE2141/5-1. A. Lu. acknowledge gratefully support of the DFG by project LU2051/1-1. A. Lu. and A. D. W. acknowledges support by DFG-TRR160, BMBF - Q.Link.X 16KIS0867, and the DFH/UFA CDFA-05-06.
\end{acknowledgments}

\bibliography{Literatur}

\end{document}